\documentclass[12pt,a4paper]{article}

\newcommand{\rx}{\mathrm{X}}
\newcommand{\te}{\tau_e}
\newcommand{\xa}{\xi_1}
\newcommand{\xb}{\xi_2}
\newcommand{\xe}{x_{0e}}
\newcommand{\de}{\Delta}
\newcommand{\la}{\lambda}
\newcommand{\pa}{\partial}
\newcommand{\del}{\delta}
\newcommand{\ms}{\mathcal{S}}
\newcommand{\mj}{\mathcal{J}}

\begin{document}

\begin{flushright}
{}
\end{flushright}
\vspace{1.8cm}

\begin{center}
 \textbf{\Large Correlators of Vertex Operators for Circular Strings \\
with Winding Numbers in $AdS_5 \times S^5$ }
\end{center}
\vspace{1.6cm}
\begin{center}
 Shijong Ryang
\end{center}

\begin{center}
\textit{Department of Physics \\ Kyoto Prefectural University of Medicine
\\ Taishogun, Kyoto 603-8334 Japan}  \par
\texttt{ryang@koto.kpu-m.ac.jp}
\end{center}
\vspace{2.8cm}
\begin{abstract}
We compute semiclassically the two-point correlator of the marginal
vertex operators describing the rigid circular spinning string state
with one large spin and one windining number in $AdS_5$ and three
large spins and three winding numbers in $S^5$. The marginality
condition and the conformal invariant expression for the two-point
correlator obtained by using an appropriate vertex operator are shown
to be associated with the diagonal and off-diagonal Virasoro
constraints respectively. We evaluate semiclassically the three-point
correlator of two heavy circular string vertex operators and one 
zero-momentum dilaton vertex operator and discuss its relation 
with the derivative of the dimension of the heavy circular string
state with respect to the string tension.
\end{abstract} 
\vspace{3cm}
\begin{flushleft}
November, 2010
\end{flushleft}

\newpage
\section{Introduction}

From the AdS/CFT correspondence \cite{JM} the correlation functions in
the $\mathcal{N}=4$ super Yang-Mills theory (SYM) can be calculated both
at weak and at strong coupling. Using the relation between the generating
functionals for the correlation functions and the AdS string partition 
functions, the three-point correlation functions
of protected operators have been derived at strong 
coupling in the supergravity approximation \cite{MV,LM}. 

The AdS/CFT correspondence suggests that each on-shell string state or 
marginal string vertex operator in $AdS_5 \times S^5$ string theory should
be associated with a local gauge-invariant operator in the $\mathcal{N}=4$
SYM theory with definite conformal dimension \cite{AP}. There has been a
prescription for computing a two-point correlation 
function of vertex operators that describe the string states with large
spins and are dual to the non-protected states with large quantum 
numbers \cite{AT}. The stationary point solution obtained by inserting
appropriate vertex operators for the two-point function should coincide
with the classical string solution carrying the same charges. 
The semiclassical computation of the two-point function by demanding
the 2d conformal invariance should lead to the same relation between
energy and spins as constructed for the corresponding classical string
solution by requiring the Virasoro constraints.

For the case that the associated string is the folded string with
large $AdS_5$ spin \cite{GKP} this connection has been demonstrated 
in the flat space limit \cite{AT}, in the global coordinates \cite{EB}
where the dependence of vertex operators on the $AdS_5$ boundary points
is ignored, and in the Poincare coordinates \cite{BT} where the locations
of vertex operators in the four-dimensional boundary space are specified
and the two-point function shows
the  scaling behavior of the strong-coupling
limit of the two-point function of single trace minimal twist 
operators in gauge theory. In ref. \cite{EB} the two-point function
associated with the string solution with large two equal spins $J_1 = J_2$
and two equal winding numbers $m_1 = m_2$  in $S^5$ has been discussed
where the vertex operator includes an additional factor expressed 
in terms of the T-dual variables.

There have been various semiclassical studies about the calculations of
correlators involving Wilson loops \cite{BC,PZ,DS,TS,MY}.

Using appropriate wave functions the two-point correlation function has
been computed \cite{JS}, where a circular heavy string state with spin $S$
in $AdS_5$ and equal spin $J (= S)$ in $S^5$ \cite{FT,AR} propagates
between two $AdS_5$ boundary points, and has been shown to produce right
scaling behavior with the scaling dimension which is determined by the
saddle point in the integraion over the modular parameter. 

This approach has been extended to computation of a three-point 
correlation function which consists of two non-BPS operators dual to 
heavy string states and one chiral primary operator dual to a
supergravity field \cite{CM}. The three-point correlator has been
derived by evaluating semiclassically a Witten diagram with a supergravity
field propagating from the $AdS_5$ 
boundary to the heavy string worldsheet.
Specially for one gauge theory Lagrangian dual to the massless dilaton
field and two non-BPS operators dual to several string configurations
such as the circular spinning string with two equal spins $S = J$ 
in $AdS_5 \times S^5$ or $J_1 = J_2$ in $S^5$ \cite{FT} and the giant 
magnon in $S^5$ \cite{HM}, the three-point correlators have been 
computed and the associated three-point couplings, 
namely, the OPE coefficients have been shown to
agree with the values predicted by 
renormalization group arguments.

On the other hand there has been a construction of a three-point 
correlator \cite{KZ} for two operators dual to the folded spinning 
string with two spins $J_1, J_2$ in $S^5$ \cite{SFT} and one primary
scalar operator dual to the supergravity massless scalar field that is 
a mixture of the metric with RR four-form \cite{LM,KR}.
The three-point correlator takes the form of a vertex operator in the
coordinate representation \cite{AP,AT} integrated over the classical 
string worldsheet.

From the vertex operator prescription \cite{AT,EB,BT} various three-point
correlators of two heavy vertex operators and one light vertex
operator have been computed \cite{RT}. The heavy vertex operator 
represents the folded spinning string 
with $AdS_5$ spin $S$ and $S^5$ spin $J$
\cite{FTT} or the rigid circular string with three spins $J_1 =J_2$ and 
$J_3$ \cite{FT} which includes the small string limit, 
while there are various choices of light vertex operators that represent
dilaton, the superconformal primary scalar state, the massive state
on the leading Regge trajectory and a special singlet string state 
on massive string levels. Further there have been constructions
of the three-point correlators between several choices of one
light vertex operator and the two heavy vertex operators representing
the circular spinning string with one $AdS_5$ spin, three different
$S^5$ spins and the corresponding winding numbers \cite{RH}. 

It is desirable to elucidate further the relations between the vertex
operators and various string solutions in $AdS_5 \times S^5$.
We are interested how to construct the vertex operators representing
the circular spinning string states with winding numbers.
Using the vertex operator prescription we will compute a two-point
correlation function of the vertex operators which describe the
rigid circular spinning string in $AdS_5 \times S^5$ \cite{AR}
carrying spin $S$ with winding number $n$ in $AdS_5$ as well as
spin $J$ with winding number $m$ in $S^5$. The extension to the circular
spinning string including three pairs of spin and winding number 
$(J_k, m_k), k=1,2,3$ in $S^5$ and the $AdS_5$ quantum numbers $(S, n)$
will be performed. We will consider a three-point correlation
function which consists of two heavy vertex operators describing this
circular multi-spin $(S, J_k)$ string  
 and one zero-momentum dilaton vertex operator.
From the three-point correlator the three-point coupling will be
constructed explicitly in terms of the relevant quantum numbers.

\section{Two-point correlator}

We consider a two-point correlation function of the string vertex 
operators which are associated with the circular spinning string solution
in the Poincare coordinates of $AdS_5 \times S^5$ with spins and winding
numbers $(S, n)$ in $AdS_5$ and $(J, m)$ in $S^5$. 

The embedding coordinates $Y_M \; (M=0, \cdots,5)$ for the Minkowski 
signature $AdS_5$ are expressed in terms of the global coordinates
$(t, \rho, \theta, \phi_1, \phi_2)$ as
\begin{eqnarray}
Y_5 + iY_0 &=& \cosh \rho e^{it}, \;\; Y_1 + iY_2 = \sinh \rho 
\cos \theta e^{i\phi_1}, \;\;Y_3 + iY_4 = \sinh \rho \sin \theta
e^{i\phi_2}, \nonumber \\
Y^MY_M &=& -Y_5^2 + Y^mY_m + Y_4^2 = -1, \hspace{1cm} 
Y^mY_m = -Y_0^2 + Y_iY_i
\label{ey}\end{eqnarray}
with $m=0,1,2,3, i=1,2,3$. These coordinates are related with the
Poincare coordinates $(z, x^m), ds^2 = z^{-2}(dz^2 + dx^mdx_m)$ 
\begin{equation}
Y_m = \frac{x_m}{z}, \hspace{1cm} Y_4 = \frac{1}{2z}(-1 + z^2 + x^mx_m),
\hspace{1cm} Y_5 = \frac{1}{2z}(1 + z^2 + x^mx_m)
\end{equation}
with $x^mx_m = -x_0^2 + x_i^2$. For $S^5$ the embedding coordinates are 
defined by 
\begin{eqnarray}
\rx_1 &\equiv& X_1 + iX_2 = \sin\gamma \cos\psi e^{i\varphi_1} =
r_1e^{i\varphi_1}, \hspace{1cm}
\rx_2 \equiv X_3 + iX_4 = \sin\gamma \sin\psi e^{i\varphi_2}= 
r_2e^{i\varphi_2}, \nonumber \\
\rx_3 &\equiv& X_5 + iX_6 = \cos\gamma e^{i\varphi_3}= 
r_3e^{i\varphi_3}, \hspace{1cm} \sum_{k=1}^3 r_k^2 = 1.
\label{ex}\end{eqnarray}

Following the procedure of ref. \cite{AT}, we rotate the worldsheet time 
$\tau$ as well as the global AdS time $t$ to the Euclidean ones 
simultaneously 
\begin{equation}
\te = i\tau, \hspace{1cm} t_e = it.
\label{eu}\end{equation}
We perform the following conformal transformation to map the Euclidean
2d cylinder $(\te, \sigma)$ into the complex plane $(\xi, \bar{\xi})$ with
two punctures at $ \xi = \xa \;(\te = \infty), \; 
\xi = \xb \;(\te = -\infty)$
\begin{equation}
e^{\te + i\sigma} = \frac{\xi - \xb}{\xi - \xa},
\label{co}\end{equation}
where $\xa$ and $\xb$ are regarded as the points where two vertex 
operators are inserted. The Euclidean rotation $t_e = it$ in (\ref{eu})
leads to the similar rotations for the time-like coordinates
\begin{equation}
Y_{0e} = iY_0, \hspace{1cm} \xe = ix_0.
\end{equation}
In general the integrated marginal vertex operator represents a string
state and is described in terms of four coordinates of a point on the
boundary of the Euclidean Poincare patch of $AdS_5$ space as
\begin{eqnarray}
V(x') &=& \int d^2\xi V( x(\xi) - x', \cdots) \nonumber \\
 &=&  \int d^2\xi [ z(\xi) + z^{-1}(\xi)(x_m(\xi) - x'_m )^2]^{-\de}
U[x_m(\xi) - x'_m, z(\xi), \rx_k(\xi) ],
\label{vu}\end{eqnarray}
where we shift $x_m = (\xe, x_i)$ by a constant vector $x'_m$.
The explicit expressions of $U$ are specified by the quantum numbers other
than the 4d dimension $\de$ of the string state in $AdS_5 \times S^5$.

Let us consider the vertex operator which describes the circular spinning
closed string state with quantum numbers like spins $(S, J)$ and
winding numbers $(n, m)$ \cite{AR}
\begin{equation}
t = \kappa \tau, \;\; \rho = \rho_0, \;\; \theta = 0, \;\;
\phi_1 = \omega \tau - n \sigma \equiv \phi, \;\;
\varphi = w\tau + m\sigma.
\label{so}\end{equation}
Its energy-spin relation is given by 
\begin{equation}
E = J + S + \frac{\la}{2J^2}(m^2J + n^2S) + \cdots.
\label{ej}\end{equation}
We propose a vertex operator 
\begin{equation}
V(a) = c \int d^2\xi [ z + z^{-1}( (\xe  - a )^2 + x_i^2 )]^{-\de}
( Y_1 + iY_2 )^S (X_1 + iX_2 )^J,
\label{va}\end{equation}
where the location of the vertex operator in the boundary is chosen by
$x'_m = (a,0,0,0)$. The expression $Y_1 + iY_2$ is described in terms of
Euclidean Poincare coordinates as
\begin{equation}
Y_1 + iY_2 = \frac{x_1 + ix_2}{z} = \frac{r}{z} e^{i\phi}.
\end{equation}
From (\ref{ex}) with $\gamma = \pi/2, \; \psi = 0$ and $\varphi_1 \equiv
\varphi$ the $S^5$ part is expressed as $X_1 + iX_2 = e^{i\varphi}$.
The proposed vertex operator shows a symmetric expression such that 
the $AdS_5$ factor $( Y_1 + iY_2 )^S$ takes the same form as the $S^5$ 
factor $(X_1 + iX_2 )^J$  including no derivatives with respect to the
worldsheet coordinates. Due to winding numbers the string solution has no
short string limit, that is, no flat space limit 
so that the expression (\ref{va}) takes a different form 
from the vertex operator in flat space which includes the derivatives.

We use this vertex operator involving the winding number dependences
implicitly through the angular coordinates to compute a 
two-point function 
\begin{equation}
< V_{\de,S,n,J,m}(a)( V_{\de,S,n,J,m}(0) )^* > \hspace{1cm}
\label{tw}\end{equation}
in large spins of order of string tension $T = \sqrt{\la}/2\pi \gg 1$.
The Euclidean continuation (\ref{eu}) of (\ref{so}) is given by
\begin{equation}
t_e = \kappa\te, \;\; \rho = \rho_0, \;\; \theta = 0, \;\;
\phi = i\omega\te + n\sigma, \;\; \varphi = iw\te - m\sigma,
\label{es}\end{equation}
where we change the signs of $\phi$ and  $\varphi$.
The corresponding embedding coordinates (\ref{ey}) are expressed as
\begin{eqnarray}
Y_5 &=& \cosh\rho_0 \cosh\kappa\te, \hspace{1cm} Y_{0e} =
 \cosh\rho_0 \sinh\kappa\te, \hspace{1cm} Y_4 = 0, \nonumber \\
Y_1 &=& \sinh\rho_0 \cosh(\omega\te - in\sigma), \hspace{1cm} 
Y_2 = i\sinh\rho_0 \sinh(\omega\te - in\sigma), \hspace{1cm} Y_3 = 0,
\end{eqnarray}
which are transformed to the Euclidean Poincare coordinates
in the form 
\begin{eqnarray}
z &=& \frac{1}{\cosh\rho_0 \cosh\kappa\te}, \hspace{1cm}
\xe = \tanh \kappa\te, \nonumber \\
x_{\pm} &\equiv& x_1 \pm ix_2 = re^{\pm i\phi} = \frac{\tanh\rho_0}
{\cosh\kappa\te} e^{\pm (in\sigma - \omega \te)},
\label{ps}\end{eqnarray}
which satisfy $z^2 + x_+x_- + \xe^2 = 1$.

For the large spin limit $\kappa \approx \omega \gg 1$
 the complex world surface
given by (\ref{ps}) approaches the boundary $z \rightarrow 0$ at
$\te \rightarrow \pm \infty$ with $\xe(\pm \infty) = \pm 1$.
By making a dilatation and a translation such that $\xe (\infty) = a,
\; \xe (-\infty) = 0$ at the boundary we have 
\begin{eqnarray}
z &=& \frac{a}{2}\frac{1}{\cosh\rho_0 \cosh\kappa\te}, \hspace{1cm}
\xe = \frac{a}{2}(\tanh \kappa\te + 1), \nonumber \\
x_{\pm} &\equiv&  x_1 \pm ix_2 = re^{\pm i\phi} = \frac{a}{2}
\frac{\tanh\rho_0}{\cosh\kappa\te} e^{\pm (in\sigma - \omega \te)}.
\label{ts}\end{eqnarray}
The end points of the Euclidean world cylinder for the transformed 
configuration are specified by
\begin{eqnarray}
\te &\rightarrow& \infty : \;\; z \rightarrow 0, \;\; \xe \rightarrow a,
\;\; r \rightarrow 0, \;\; x_+ \rightarrow 0, \;\; 
x_- \rightarrow a\tanh \rho_0e^{-in\sigma}, \nonumber \\
\te &\rightarrow& -\infty : \;\; z \rightarrow 0, \;\; \xe \rightarrow 0,
\;\; r \rightarrow 0, \;\; x_+ \rightarrow a\tanh \rho_0e^{in\sigma},\;\;
x_- \rightarrow 0.
\end{eqnarray} 

When we compute semiclassically the two-point correlator (\ref{tw}),
the Euclidean action accompanied with the vertex contributions is
given by
\begin{eqnarray}
A_e &=& \frac{\sqrt{\la}}{\pi}\int d^2\xi \left[ \frac{1}{z^2}
( \pa z\bar{\pa}z + \pa \xe \bar{\pa}\xe + \pa r\bar{\pa}r
+ r^2\pa \phi\bar{\pa}\phi ) + \pa \varphi\bar{\pa}\varphi \right]
\nonumber \\
 &-& \de \int d^2\xi \left[ \del^2(\xi - \xa) \ln 
\frac{z}{z^2 + r^2 + (\xe - a)^2} + \del^2(\xi - \xb) \ln 
\frac{z}{z^2 + r^2 + \xe^2} \right] \nonumber \\
&-& S \int d^2\xi \left[ \del^2(\xi - \xa) \ln \frac{re^{i\phi}}{z}
+ \del^2(\xi - \xb) \ln \frac{re^{-i\phi}}{z} \right] \nonumber \\
&-& J \int d^2\xi [ \del^2(\xi - \xa) \ln e^{i\varphi}
+ \del^2(\xi - \xb) \ln e^{-i\varphi} ].
\label{ae}\end{eqnarray}
We will show that the transformed complex solution (\ref{ts})
with $\varphi$ in (\ref{es}) becomes the stationary 
trajectory in the presence of the
vertex operators as source terms.
The equation of motion for $\phi$ is obtained by
\begin{equation}
\pa\left( \frac{r^2}{z^2} \bar{\pa}\phi \right) + 
\bar{\pa}\left( \frac{r^2}{z^2} \pa\phi \right) = - \frac{i\pi S}
{\sqrt{\la}} [ \del^2(\xi - \xa)  - \del^2(\xi - \xb)].
\label{ph}\end{equation}
The inversion of the conformal transformation (\ref{co}) is given by
\begin{equation}
\te = \frac{1}{2} \ln \frac{(\xi - \xb)(\bar{\xi} - \bar{\xb})}
{(\xi - \xa)(\bar{\xi} - \bar{\xa})}, \hspace{1cm}
\sigma = \frac{1}{2i} \ln \frac{(\xi - \xb)(\bar{\xi} - \bar{\xa})}
{(\bar{\xi} - \bar{\xb})(\xi - \xa)},
\end{equation}
which leads to
\begin{equation}
(\pa\bar{\pa} + \bar{\pa}\pa)\te = \pi [ \del^2(\xi - \xb)  - 
\del^2(\xi - \xa)], \hspace{1cm}
(\pa\bar{\pa} + \bar{\pa}\pa)\sigma = 0.
\label{te}\end{equation}
Taking account of (\ref{te}) and (\ref{ts}) we see that the equation 
(\ref{ph}) is satisfied only when $S$ is given by
\begin{equation}
S = \omega \sinh^2\rho_0 \sqrt{\la}.
\label{sh}\end{equation}

We turn to the equation of motion for $\xe$ 
\begin{equation}
\pa \frac{\bar{\pa}\xe}{z^2} + \bar{\pa} \frac{\pa\xe}{z^2} =
\frac{2\pi\de}{\sqrt{\la}} \left[ \frac{\xe - a}{z^2 + r^2 + (\xe - a)^2}
\del^2(\xi - \xa) + \frac{\xe}{z^2 + r^2 + \xe^2}\del^2(\xi - \xb)\right],
\end{equation}
which becomes through (\ref{ts}) 
\begin{equation}
\kappa \cosh^2\rho_0 (\pa\bar{\pa} + \bar{\pa}\pa)\te = 
\frac{\pi\de}{\sqrt{\la}} [ \del^2(\xi - \xb) - \del^2(\xi - \xa)].
\label{xo}\end{equation}
Owing to (\ref{te}) this equation leads to
\begin{equation}
\de = \kappa \cosh^2\rho_0 \sqrt{\la}.
\label{dk}\end{equation}
The equation of motion for $r$ is expressed as 
\begin{eqnarray}
r\pa \frac{\bar{\pa}r}{z^2} + r\bar{\pa} \frac{\pa r}{z^2} 
- 2\frac{r^2}{z^2}\pa \phi \bar{\pa}\phi = - \frac{\pi S}{\sqrt{\la}}
[ \del^2(\xi - \xa) + \del^2(\xi - \xb)] \nonumber \\
+ \frac{2\pi\de}{\sqrt{\la}} \left[ \frac{r^2}{z^2 + r^2 + (\xe - a)^2}
\del^2(\xi - \xa) + \frac{r^2}{z^2 + r^2 + \xe^2}
\del^2(\xi - \xb)\right].
\label{rr}\end{eqnarray}
Since $\te \rightarrow \pm\infty$ corresponds to $\xi \rightarrow 
\xi_{1,2}$ it reads
\begin{eqnarray}
- \kappa [ 2\kappa \pa\te \bar{\pa}\te + \tanh\kappa\te 
(\pa\bar{\pa} + \bar{\pa}\pa)\te ] - 2(n^2 - \omega^2)
\pa\te \bar{\pa}\te \nonumber \\
= \pi ( 2\kappa - \omega )[\del^2(\xi - \xa) + \del^2(\xi - \xb)],
\label{re}\end{eqnarray}
where we use $\pa\sigma \bar{\pa}\sigma = \pa\te \bar{\pa}\te$,
$\pa\sigma \bar{\pa}\te +  \bar{\pa}\sigma \pa\te = 0$, (\ref{sh})
and (\ref{dk}).  
In view of the non-singular $\pa\te \bar{\pa}\te$ terms in (\ref{re})
we have 
\begin{equation}
\omega^2 = \kappa^2 + n^2
\label{kn}\end{equation}
and the remaining singular terms provide
\begin{equation}
\kappa [\tanh\kappa \te  \del^2(\xi - \xa) - \tanh\kappa \te
 \del^2(\xi - \xb) ] =  (2\kappa - \omega )[\del^2(\xi - \xa) + 
\del^2(\xi - \xb)].
\label{kt}\end{equation}
Owing to  
\begin{equation}
\tanh\kappa \te|_{\xi \rightarrow \xa} = 1, \hspace{1cm}
\tanh\kappa \te|_{\xi \rightarrow \xb} = -1
\label{tt}\end{equation}
the equation (\ref{kt}) is satisfied  to the leading order in the large
spin limit $\kappa \approx \omega \gg 1$.

Combining the equation of motion for $z$ and that for $r$ 
(\ref{rr}) we have
\begin{equation}
\frac{1}{2}\pa \frac{\bar{\pa}( z^2 + r^2)}{z^2} + \frac{1}{2}\bar{\pa}
\frac{\pa ( z^2 + r^2)}{z^2} 
+ \frac{2}{z^2}\pa \xe \bar{\pa}\xe =  \frac{\pi \de}{\sqrt{\la}}
[ \del^2(\xi - \xa) + \del^2(\xi - \xb)],
\end{equation}
which takes the following compact form
\begin{equation}
\kappa \cosh^2\rho_0(\pa\bar{\pa} + \bar{\pa}\pa)\te = 
-\frac{\pi \de}{\sqrt{\la}} \left[ \frac{\del^2(\xi - \xa)}
{\tanh\kappa \te} + \frac{\del^2(\xi - \xb)}{\tanh\kappa \te} \right].
\end{equation}
Because of (\ref{tt}) this equation coincides with the equation of  
motion for $\xe$ (\ref{xo}). The remaining equation of motion for 
$\varphi$ reads
\begin{equation}
(\pa\bar{\pa} + \bar{\pa}\pa)\varphi = -\frac{i\pi J}{\sqrt{\la}}
[ \del^2(\xi - \xa) - \del^2(\xi - \xb)].
\end{equation}
It holds  if $J$ is given by
\begin{equation}
J = w\sqrt{\la}.
\label{jw}\end{equation}

Now the two-point correlation function can be calculated semiclassically
by evaluating the string action with the source terms on the stationary
string trajectory. It is convenient to go back to the Euclidean 2d 
cylinder $(\te, \sigma)$ coordinates for computing 
the string action in (\ref{ae})
\begin{eqnarray}
A_{str} &=& \frac{\sqrt{\la}}{4\pi} \int_{\tau_{e,-\infty}}
^{\tau_{e,\infty}} d\te \int_0^{2\pi}d\sigma \biggl[ \frac{1}{z^2}
\biggl( (\pa_{\te}z)^2 + (\pa_{\te}\xe)^2 + (\pa_{\te}r)^2
+ r^2( (\pa_{\te}\phi)^2 + (\pa_{\sigma}\phi)^2 ) \biggr) \nonumber \\
&+& (\pa_{\te}\varphi)^2 + (\pa_{\sigma}\varphi)^2 
\biggr]
\label{as}\end{eqnarray}
with
\begin{equation}
\tau_{e,\pm\infty} = \frac{1}{2}( \ln |\xi - \xb |^2 - \ln |\xi - \xa |^2)
|_{\xi \rightarrow \xi_{1,2} } .
\end{equation}
The substitution of the stationary solution 
into (\ref{as}) yields
\begin{equation}
A_{str} = \frac{\sqrt{\la}}{2} [ \kappa^2\cosh^2\rho_0 + 
( n^2 - \omega^2 )\sinh^2\rho_0 + m^2 - w^2 ](\tau_{e,\infty}
- \tau_{e,-\infty}).
\end{equation}
We neglect the one-point function divergence $(\sim \ln 0)$ to obtain
\begin{equation}
A_{str} = \frac{\sqrt{\la}}{2} [ \kappa^2\cosh^2\rho_0 + 
( n^2 - \omega^2 )\sinh^2\rho_0 + m^2 - w^2 ]\ln |\xa - \xb |^2,
\end{equation}
while the source terms in (\ref{ae}) are also evaluated using the
delta-function as
\begin{eqnarray}
A_{sour} &=&  2\de \ln a + ( -\de\kappa + S\omega + Jw )
\ln|\xa - \xb |^2 \nonumber \\
 &+& ( Jm - Sn ) \ln \frac{\xa - \xb}{\bar{\xa} - \bar{\xb}}.
\end{eqnarray}
When two spins $S, \; J$ obey a symmetric relation
\begin{equation}
Sn = Jm,
\label{sj}\end{equation}
which is associated with the symmetric form of the 
vertex operator (\ref{va}), we derive a 2d conformal invariant 
expression for the two-point correlation function
\begin{eqnarray}
< V_{\de,S,n,J,m}(a)(V_{\de,S,n,J,m}(0))^* > \approx \int d^2\xa d^2\xb
e^{- A_{str} - A_{sour}} \nonumber \\
\approx \frac{1}{a^{2\de}} \int d^2\xa d^2\xb \frac{1}
{|\xa - \xb|^{\sqrt{\la}[ -\frac{\de\kappa}{\sqrt{\la}} + 
\sinh^2\rho_0( \omega^2 + n^2 ) + w^2 + m^2 ] }  }.
\label{pd}\end{eqnarray}

This expression shows the scaling $\sim a^{-2\de}$ which is characterized
by the scaling dimension $\de$ of the vertex operator and is expected from
the 4d conformal invariance of string action under the rescaling 
$z \rightarrow az, x_m \rightarrow ax_m$. From the marginality condition
of vertex operator the two-point correlator (\ref{pd}) should take a 
behavior $|\xa - \xb|^{-4}$ which in the large spin leads to 
\begin{equation}
\frac{\kappa\de}{\sqrt{\la}} - \sinh^2\rho_0( \omega^2 + n^2 )
- w^2 - m^2 = 0.
\end{equation}
Using (\ref{sh}), (\ref{dk}), (\ref{kn}) we eliminate $n$ and $\rho_0$ 
to obtain
\begin{equation}
\frac{2\kappa\de}{\sqrt{\la}} - 2\omega \ms - \kappa^2 = \mj^2 + m^2
\label{jm}\end{equation}
with $S = \ms \sqrt{\la}, \; J = \mj \sqrt{\la}$.

In ref. \cite{AR} the circular two-spin $(S,J)$ string solution with the
corresponding winding numbers $(n,m)$ was constructed to be specified
by the same relations between the relevant parameters as (\ref{sh}),
(\ref{dk}), (\ref{kn}), (\ref{jw}) where the off-diagonal 
Virasoro constraint gives the same relation as (\ref{sj}) and
the diagonal Virasoro constraint is presented by
\begin{equation}
2\kappa \mathcal{E} - 2\omega \ms - \kappa^2 = 2\sqrt{m^2 + \nu^2} \mj
- \nu^2 
\label{js}\end{equation}
with $\mj = \sqrt{m^2 + \nu^2}$. 

The marginality condition (\ref{jm}) becomes the same expression as
(\ref{js}) when the parameter $\nu$ is eliminated and the dimension $\de$
is identified with the string energy $E = \mathcal{E}\sqrt{\la}$.
Therefore the dimension $\de$ is determined as (\ref{ej}).
We observe that the stationary trajectory of path integral representing
the two-point function of vertex operators is provided by conformally
mapped Euclidean continuation of the circular two-spin $(S,J)$ 
string solution \cite{AR}, whose energy-spin relation (\ref{ej}) is 
derived from the Virasoro constraints for the worldsheet conformal
invariance. Thus the two-point correlation function (\ref{pd}) shows 
the right scaling behavior with the dimension $\de$ being the 
corresponding string energy (\ref{ej}).

Now we consider the two-point correlator of the vertex operators
representing the rigid circular spinning string which has spin $S$ with
winding number $n$ in $AdS_5$ and three spins $J_k, \; k=1,2,3$ with
the corresponding winding numbers $m_k$ in $S^5$. We take the
following vertex operator
\begin{equation}
V(a) = c \int d^2\xi [ z + z^{-1}( (\xe  - a )^2 + x_i^2 )]^{-\de}
( Y_1 + iY_2 )^S \rx_1^{J_1}\rx_2^{J_2} \rx_3^{J_3}.
\label{vx}\end{equation}
The Euclidean action with the source terms from the vertex operators
is also expressed as
\begin{eqnarray}
A_e &=& \frac{\sqrt{\la}}{\pi}\int d^2\xi \left[ \frac{1}{z^2}
( \pa z\bar{\pa}z + \pa \xe \bar{\pa}\xe + \pa r\bar{\pa}r
+ r^2\pa \phi\bar{\pa}\phi ) + 
\frac{1}{2}\sum_{k=1}^3 (\pa \rx_k \bar{\pa}\bar{\rx}_k +
\bar{\pa} \rx_k \pa\bar{\rx}_k ) \right] \nonumber \\
 &-& \de \int d^2\xi \left[ \del^2(\xi - \xa) \ln 
\frac{z}{z^2 + r^2 + (\xe - a)^2} + \del^2(\xi - \xb) \ln 
\frac{z}{z^2 + r^2 + \xe^2} \right] \nonumber \\
&-& S \int d^2\xi \left[ \del^2(\xi - \xa) \ln \frac{re^{i\phi}}{z}
+ \del^2(\xi - \xb) \ln \frac{re^{-i\phi}}{z} \right] \nonumber \\
&-& \sum_{k=1}^3 J_k \int d^2\xi [ \del^2(\xi - \xa) \ln r_ke^{i\varphi_k}
+ \del^2(\xi - \xb) \ln r_ke^{-i\varphi_k} ].
\end{eqnarray}
This circular spinning string \cite{AR} has  the following Euclidean
continuation 
\begin{eqnarray}
t_e &=& \kappa\te, \;\; \rho = \rho_0, \;\; \theta = 0, \;\;
\phi = i\omega\te + n\sigma, \nonumber \\
r_k &=& const, \;\; \varphi_k = iw_k\te - m_k\sigma.
\label{ko}\end{eqnarray}

The equations of motion for $\varphi_k$ are given by
\begin{equation}
r_k^2( \pa\bar{\pa} + \bar{\pa}\pa)\varphi_k = -
\frac{i\pi J_k}{\sqrt{\la}}[ \del^2(\xi - \xa) - \del^2(\xi - \xb)],
\;\; k = 1, 2, 3,
\end{equation}
which yield
\begin{equation}
J_k = w_k r_k^2 \sqrt{\la}.
\end{equation}
The equations of motion for $r_k$, or for $\psi$ and $\gamma$, have
additional contributions to the singular part. Here we assume
that they are still satisfied by constant $r_k$. 
The string action  evaluated on this multi-spin stationary solution is
expressed as
\begin{equation}
A_{str} = \frac{\sqrt{\la}}{2} [ \kappa^2\cosh^2\rho_0 + 
( n^2 - \omega^2 )\sinh^2\rho_0 + \sum_{k=1}^3 r_k^2(m_k^2 - w_k^2) ]
\ln |\xa - \xb |^2,
\end{equation}
where the one-point function divergence is ignored.
The source terms are computed by
\begin{eqnarray}
A_{sour} &=& 2\de \ln a + ( -\de\kappa + S\omega 
+ \sum_{k=1}^3 J_kw_k )\ln|\xa - \xb |^2 \nonumber \\
 &+& ( \sum_{k=1}^3 J_km_k - Sn ) \ln \frac{\xa - \xb}{\bar{\xa} 
- \bar{\xb}}.
\end{eqnarray}
When there is a relation
\begin{equation}
Sn = \sum_{k=1}^3 J_km_k, 
\label{nm}\end{equation}
we obtain the following two-point function that is consistent
with 2d conformal symmetry
\begin{eqnarray}
&<& V_{\de,S,n,J_k,m_k}(a)(V_{\de,S,n,J_k,m_k}(0))^* \;\;> 
\nonumber \\
&\approx& \frac{1}{a^{2\de}} \int d^2\xa d^2\xb \frac{1}
{|\xa - \xb|^{\sqrt{\la}[ -\frac{\de\kappa}{\sqrt{\la}} + 
\sinh^2\rho_0( \omega^2 + n^2 ) + \sum_k r_k^2
(w_k^2 + m_k^2) ] }  }.
\end{eqnarray}
The marginality condition of the vertex operator implies
\begin{equation}
\frac{2\kappa \de}{\sqrt{\la}} - 2\omega \ms - \kappa^2 = 
\sum_{k=1}^3r_k^2 (w_k^2 + m_k^2 ).
\label{vd}\end{equation}

The circular multi-spin $(S, J_k)$ string solution with winding numbers
$(n, m_k), \; k=1,2,3$ was constructed \cite{AR} from the off-diagonal 
Virasoro constraint that agrees with (\ref{nm}) and 
the diagonal Virasoro constraint
\begin{equation}
2\kappa \mathcal{E} - 2\omega \ms - \kappa^2 = 2\sum_{k=1}^3
\sqrt{m_k^2 + \nu^2} \mj_k - \nu^2 
\label{mj}\end{equation}
with $\mj_k = r_k^2w_k, \; w_k = \sqrt{m_k^2 + \nu^2}$.
We use an identity $\nu^2 = \sum_k r_k^2\nu^2 
= \sum_k r_k^2 (w_k^2 - m_k^2)$
to see that the two relations (\ref{vd}) and (\ref{mj}) coincide
for $\de = \mathcal{E}\sqrt{\la}$.  These agreements imply that
the symmetric vertex operator (\ref{vx}) is a natural candidate
describing the circular spinning string with quantum numbers like
spins $(S, J_k)$ and winding numbers $(n, m_k)$  in $AdS_5 \times S^5$.

\section{Three-point correlator}

Let us consider the three-point correlation function of two heavy vertex
operators $V_H$ representing a large spin string state and one light 
vertex operator $V_L$ representing a massless string
state \cite{RT} 
\begin{equation}
< V_{H_1}(x_1)V_{H_2}(x_2)V_L(x_3) >.
\label{hl}\end{equation}
We choose  $V_{\de,S,n,J_k,m_k}(x_1)(V_{\de,S,n,J_k,m_k}(x_2))^* $
as the product of two heavy vertex operators $V_{H_1}(x_1)V_{H_2}(x_2)$,
where the vertex operators are located on the boundary points
$x_1^m = (1,0,0,0)$ and $x_2^m = (-1,0,0,0)$ so that we use the
multi-spin $(S,J_k)$ string Euclidean solution (\ref{ps}) with (\ref{ko})
in stead of (\ref{ts}) to evaluate the three-point correlator (\ref{hl})
semiclassically. Fixing the location of $V_L(x_3)$ as 
$x_3^m = (0,0,0,0)$ we express the light vertex operator  in the same form
as (\ref{vu})
\begin{equation}
V_L(0) = \int d^2\xi (Y_+)^{-\de_L}U[x_m(\xi), z(\xi), \rx_k(\xi) ]
\end{equation}
with $Y_+ = Y_4 + Y_5 = z(\xi) + z^{-1}(\xi)x_m^2(\xi)$. 
The three-point correlator (\ref{hl}) is estimated  semiclassically by 
integrating over the worldsheet of the heavy string  as
\begin{eqnarray}
< V_{H_1}(x_1)V_{H_2}(x_2)V_L(0) > &\approx& \int d^2\xi 
(Y_+^{cl})^{-\de_L}U[x_m^{cl}(\xi), z^{cl}(\xi), \rx_k^{cl}(\xi) ]
\nonumber \\
 &=& \int d^2\xi (z^{cl})^{\de_L}
U[x_m^{cl}(\xi), z^{cl}(\xi), \rx_k^{cl}(\xi) ],
\end{eqnarray}
where the heavy string configuration (\ref{ps}) satisfies 
$z^2 + x_m^2 = 1$ and the superscript represents 
the stationary point solution for the 
two-point function of two heavy vertex operators.
The structure coefficient $C_{123}$ is obtained by
\begin{equation}
C_{123} = \frac{< V_{H_1}(x_1)V_{H_2}(x_2)V_L(0) >}
{< V_{H_1}(x_1)V_{H_2}(x_2)>} = c_{\de_L}\int d^2\xi (z^{cl})^{\de_L}
U[x_m^{cl}(\xi), z^{cl}(\xi), \rx_k^{cl}(\xi) ],
\end{equation}
where $c_{\de_L}$ depends on the normalization of the light vertex 
operator.

As  the unintegrated light vertex operator for $V_L$ 
we choose the massless string vertex 
operator dual to the dilaton field with $\de_L \equiv \de$
\begin{equation}
V_j^{(dil)} = (Y_+)^{-\de}(X_3)^j\sqrt{\la} \left(\pa Y_M \bar{\pa}Y^M +
\frac{1}{2}\sum_{k=1}^3 (\pa \rx_k \bar{\pa}\bar{\rx}_k 
+ \bar{\pa} \rx_k \pa\bar{\rx}_k ) \right),
\end{equation}
which represents a highest weight state of $SO(2,4) \times SO(6)$ and is
proportional to the string Lagrangian including the string tension
$\sqrt{\la}/2\pi$. The marginality condition for $V_j^{(dil)}$
gives $\de = 4 + j$ to the leading order in the large $\sqrt{\la}$
expansion. The corresponding dual gauge theory operator is
$tr(F_{mn}^2Z^j + \cdots )$ which becomes the SYM Lagrangian when the
KK momentum $j$ of the dilaton field is zero.

Here we devote ourselves to the $j = 0$ case. The three-point function
coefficient $C_{123}$ is given by
\begin{equation}
C_{123} = \frac{c_{\de}\sqrt{\la}}{4} \int_{-\infty}^{\infty}d\te
 \int_0^{2\pi} d\sigma z^4 \sum_{a=\te,\sigma}
\left[ \frac{(\pa_a x_m)^2 + (\pa_a z)^2 }{z^2} +  
\frac{1}{2}\sum_{k=1}^3 (\pa_a \rx_k
 \pa_a\bar{\rx}_k + \pa_a \rx_k \pa_a\bar{\rx}_k ) \right]. 
\label{cz}\end{equation}
We substitute the stationary multi-spin $(S,J_k)$ string 
Euclidean solution (\ref{ps}) with (\ref{ko}) into (\ref{cz}) to have
\begin{equation}
C_{123} = \frac{c_{\de}\sqrt{\la}}{4} \int_{-\infty}^{\infty}d\te
 \int_0^{2\pi} d\sigma \frac{\kappa^2 + \sum_k r_k^2
( m_k^2 - w_k^2 )}{( \cosh\rho_0 \cosh\kappa\te )^4},
\label{in}\end{equation}
whose integration over $\te$ leads to a finite value
\begin{equation}
C_{123} = \frac{2c_{\de}\pi}{3}\frac{1}{\cosh^4\rho_0}
\frac{(\kappa^2 - \nu^2 )\sqrt{\la}}{\kappa}.
\label{ck}\end{equation}

The diagonal Virasoro constraint (\ref{mj}) together with
\begin{equation}
\frac{\mathcal{E}}{\kappa} - \frac{\ms}{\omega} = 1
\label{oe}\end{equation}
can be expressed as 
\begin{equation}
\kappa^2 + \nu^2 = \frac{2n^2}{\sqrt{\kappa^2 + n^2}}\ms +
2\sum_k \sqrt{\nu^2 + m_k^2} \mj_k.
\label{nn}\end{equation}
The equation 
\begin{equation}
\sum_k r_k^2 = \sum_k \frac{\mj_k}{\sqrt{\nu^2 + m_k^2}} = 1
\label{sr}\end{equation}
and (\ref{nn}) are solved by using the large $\mj$ expansion
with $\mj \equiv \sum_k \mj_k$
\begin{eqnarray}
\kappa &=& \mj + \frac{1}{2\mj^2}( \sum_k m_k^2\mj_k + 2n^2\ms ) 
\nonumber \\
 &-& \frac{1}{8\mj^5}( \mj\sum_k m_k^4\mj_k + 4n^4\ms \mj +
8n^2 \ms \sum_k m_k^2\mj_k + 12n^4 \ms^2 ) + \cdots, 
\nonumber \\
\nu &=& \mj - \frac{1}{2\mj^2} \sum_k m_k^2\mj_k 
+ \frac{1}{8\mj^5}[ 3\mj\sum_k m_k^4\mj_k - 4(\sum_k m_k^2\mj_k)^2 ] 
 + \cdots.
\label{nt}\end{eqnarray}
Combining together we have the energy-spin relation in the $\la/J^2$
expansion with total spin $J = \mj\sqrt{{\la}}$
\begin{eqnarray}
E &=& J + S + \frac{\la}{2J^2}(\sum_k m_k^2J_k + n^2S ) \nonumber \\
&-& \frac{\la^2}{8J^5}( J\sum_k m_k^4J_k + n^4SJ +
4n^2 S\sum_k m_k^2J_k + 4n^4S^2 ) + \cdots.
\label{ee}\end{eqnarray}

The substitution of the expansion (\ref{nt}) into (\ref{ck}) yields
\begin{eqnarray}
C_{123} &\approx& \frac{4c_{\de}\pi}{3}\frac{1}{\cosh^2\rho_0}
\biggl[ \frac{\la}{J^2}(\sum_k m_k^2J_k + n^2S ) \nonumber \\
&-& \frac{\la^2}{2J^5}( J\sum_k m_k^4J_k + n^4SJ + 4n^2 S\sum_k m_k^2J_k
+ 4n^4S^2 ) + \cdots \biggr].
\label{ct}\end{eqnarray}
The factor $\cosh^2\rho_0 = 1 + \ms/\sqrt{\kappa^2 + n^2}$ is expressed
in the $\la/J^2$ expansion by using (\ref{nt}) as
\begin{equation}
\cosh^2\rho_0 = 1 + \frac{S}{J} - \frac{\la}{2J^3}\left( n^2S 
+ \frac{S}{J}( \sum_k m_k^2J_k + 2n^2S ) \right) + \cdots.
\end{equation}
Making the $\la$-derivative of the heavy state dimension $\de$, that is,
the string energy $E$ (\ref{ee}) under the fixed quantum numbers like
$S, n, J_k, m_k$ we have
\begin{eqnarray}
\sqrt{\la} \frac{\pa\de}{\pa \sqrt{\la}}  &=& \frac{\la}{J^2}
(\sum_k m_k^2J_k + n^2S ) \nonumber \\
&-& \frac{\la^2}{2J^5}( J\sum_k m_k^4J_k + n^4SJ + 4n^2 S\sum_k m_k^2J_k
+ 4n^4S^2 ) + \cdots .
\label{lj}\end{eqnarray} 
We observe that the resulting expression (\ref{lj}) is contained in the
three-point coupling $C_{123}$ (\ref{ct}).

If we use the string Lagrangian    
instead of the zero-momentum dilaton vertex operator 
which includes a $z^4$ factor, the three-point
coupling $C_{123}$ has no $1/\cosh^4\rho_0$ factor but the integration
over $\te$ needs some IR cutoff. Here in the expression (\ref{in})
the IR divergence is regularized by the $z^4$ factor and we 
show that the three-point coupling $C_{123}$ includes 
$\sqrt{\la} \frac{\pa}{\pa \sqrt{\la}} \de$ by using the
large $\mj$ expansion. This kind of proportionality was presented 
by using the vertex operator procedure, where the light state is 
described by the zero-momentum dilaton vertex operator and
the associated heavy state is given by the circular string solution
with two equal spins $J_1=J_2$ \cite{RT} or further two different
spins $J_1 \neq J_2$ \cite{RH}.
The relations between the derivative of the heavy state dimensions 
over the 't Hooft coupling $\la$ and
the three-point couplings for the gauge theory Lagrangian 
and two non-BPS operators dual to heavy string states were studied 
at strong coupling \cite{CM} by using the wave function procedure 
of ref. \cite{JS} and the equality between the gauge theory generating
functional for the correlators and the string partition function, 
 where the associated heavy states are provided by  the circular
string solution with two equal spins $S = J$ or $J_1=J_2$,
and the giant magnon in $S^5$. 
 
Let us make the $\sqrt{\la}$-derivative of the following 
expression obtained from (\ref{oe})
\begin{equation}
\de = \kappa \sqrt{\la} + \frac{\kappa S}{\sqrt{\kappa^2 + n^2}}
\label{dl}\end{equation}
keeping the quantum numbers
$S$ and $n$ fixed to have 
\begin{equation}
\frac{\pa\de}{\pa \sqrt{\la}} = \kappa + \frac{\pa\kappa}{\pa \sqrt{\la}}
\sqrt{\la}\left( 1 + \frac{n^2S}{\sqrt{\la}( \kappa^2 + n^2)^{3/2}}
\right),
\label{ld}\end{equation} 
where $\kappa$ is regarded as a function of $\la$. The equation (\ref{nn})
is rewritten by
\begin{equation}
\kappa^2 + \nu^2 = \frac{2n^2}{\sqrt{\la(\kappa^2 + n^2)}}S +
2 \sum_k \sqrt{ \frac{\nu^2 + m_k^2 }{\la} }J_k,
\label{lk}\end{equation}
whose derivative over $\sqrt{\la}$ under the fixed quantum numbers
yields
\begin{eqnarray}
\kappa \frac{\pa\kappa}{\pa \sqrt{\la}} \left( 1 + 
\frac{n^2S}{\sqrt{\la}( \kappa^2 + n^2)^{3/2}} \right) + 
\nu \frac{\pa\nu}{\pa \sqrt{\la}} \left( 1 - \frac{1}{\sqrt{\la}}
\sum_k \frac{J_k}{\sqrt{\nu^2 + m_k^2}} \right) \nonumber \\
= - \frac{n^2S}{\la \sqrt{ \kappa^2 + n^2}} - \frac{1}{\la}
\sum_k \sqrt{\nu^2 + m_k^2} J_k,
\label{di}\end{eqnarray}
where $\pa \nu/\pa \sqrt{\la}$ can be computed from (\ref{sr})
but its coefficient factor already vanishes.
Combining (\ref{di}) with (\ref{ld}) we have
\begin{equation}
\sqrt{\la}\frac{\pa\de }{\pa \sqrt{\la}}= \frac{1}{\kappa}
\left[ \kappa^2\sqrt{\la} - \left( \frac{n^2S}{\sqrt{ \kappa^2 + n^2}} +
\sum_k \sqrt{\nu^2 + m_k^2} J_k \right) \right],
\end{equation}
for which we use (\ref{lk}) again to derive 
\begin{equation}
\sqrt{\la}\frac{\pa\de}{\pa \sqrt{\la}} = \sqrt{\la}
\frac{\kappa^2 - \nu^2}{2\kappa}.
\end{equation}
Thus without resort to the $1/\mj$ expansion we show exactly that
$\sqrt{\la}\frac{\pa}{\pa \sqrt{\la}}\de$ is contained in the
relevant three-point coupling (\ref{ck}).

\section{Conclusion}

Based on the vertex operator prescription \cite{AT,EB,BT} we have
constructed the two-point correlation function of the vertex operators
representing the circular spinning string solution \cite{AR}
with spins $(S, J)$ or $(S, J_k), \; k=1,2,3$ as well as
winding numbers $(n, m)$ or $(n, m_k)$ in $AdS_5 \times S^5$.
We have chosen an appropriate corresponding vertex operator and shown
that the analytically continued version of this multi-spin string 
solution mapped onto complex plane is the same as the stationary point 
solution for the two-point function by demonstrating that it indeed 
solves the relevant equations of motion 
on the complex plane with the
delta-function sources at two insertion points of vertex operators.

We have observed that the marginality condition of the vertex 
operator for the 2d scaling behavior of the semiclassically evaluated
two-point function yields the same relation among energy (dimension),
spins and winding numbers as is obtained from the diagonal Virasoro
constraint, while the requirement for the two-point function to have
the 2d conformal invariant expression gives the same 
relation between the spin winding-number pairs $(S, n)$ and $(J_k, m_k)$
as follows from the off-diagonal Virasoro constraint.
These two coincidences confirm the validity of the proposed symmetric
vertex operator with no derivatives over the worldsheet coordinates. 

We have computed the three-point coupling from evaluating the 
three-point correlation function of two heavy string vertex operators
representing the circular multi-spin $(S, J_k)$ string with winding
numbers $(n, m_k)$ and one zero-momentum dilaton vertex operator
by using the stationary string surface saturating the two-point
correlation function of two heavy string vertex operators.
By using the $1/\mj$ expansion we have demonstrated that the 
three-point coupling contains
a factor representing  the derivative of the dimension of the
heavy string state with respect to the string tension . We have further
confirmed this relation exactly by manipulating various relations
between the relevant parameters and the quantum numbers of the rigid
circular multi-spin string solution.

\end{document}